\documentclass[twocolumn,a4paper]{webofc}
\usepackage[varg]{txfonts}   % Web of Conferences font
\usepackage{subcaption}
\begin{document}
\title{Cosmic ray acceleration to ultrahigh energy in radio galaxies}
\def\mnras{MNRAS, }
\def\araa{ARA\&A, }
\def\apj{ApJ, }
\def\aap{A\&A, }
\def\apjl{ApJL, }
\def\apj{ApJ, }
\def\mnras{MNRAS, }
\def\aj{AJ, }                      % "Astron. J."
\def\apjs{ApJS, }                  % "Astrophys. J. Suppl. Ser."
\def\pasp{PASP, }                  % "Publ. Astron. Soc. Pac."
\def\pasj{PASJ, }
\def\prd{Phys. Rev. D, }
\def\jcap{JCAP, }
\def\physscr{Physica Scripta, }
\def\prl{Phys. Rev. Lett., } 

\author{\firstname{James H.} \lastname{Matthews}\inst{1}\fnsep\thanks{\email{james.matthews@physics.ox.ac.uk}} \and
        \firstname{Anthony R.} \lastname{Bell}\inst{2}\fnsep \and
        \firstname{Anabella T.} \lastname{Araudo}\inst{3,4}\fnsep \and 
        \firstname{Katherine M.} \lastname{Blundell}\inst{1}\fnsep
        % etc.
}

\institute{University of Oxford, Astrophysics, Keble Road, Oxford, OX1 3RH, UK
\and
          University of Oxford, Clarendon Laboratory, Parks Road, Oxford OX1 3PU, UK
\and
          Astronomical Institute, Academy of Sciences of the Czech Republic, Bo\v{c}n\'{\i} II 1401, CZ-141\,00 Prague, Czech Republic
\and 
          ELI Beamlines, Institute of Physics, Czech Academy of Sciences, 25241 Doln{\'\i} B\v re\v zany, Czech Republic
          }

\abstract{%
  The origin of ultrahigh energy cosmic rays (UHECRs) is an open question. In this proceeding, we first review the general physical requirements that a source must meet for acceleration to 10-100 EeV, including the consideration that the shock is not highly relativistic. We show that shocks in the backflows of radio galaxies can meet these requirements. We discuss a model in which giant-lobed radio galaxies such as Centaurus A and Fornax A act as slowly-leaking UHECR reservoirs, with the UHECRs being accelerated during a more powerful past episode. We also show that Centaurus A, Fornax A and other radio galaxies may explain the observed anisotropies in data from the Pierre Auger Observatory, before examining some of the difficulties in associating UHECR anisotropies with astrophysical sources. 
}
\maketitle
\section{Introduction}
\label{intro}
The origin of ultrahigh energy cosmic rays (UHECRs) has been a mystery for decades. Various subclasses of active galactic nuclei (AGN) have been frequently discussed as possible sources \citep[e.g.][]{hillas_origin_1984,rachen1993,massaglia_role_2007,hardcastle_which_2010}, but associations with individual sources have proven difficult due to the rarity of events at ultrahigh energy and the complicating effect of intervening magnetic fields. However, recent observational results from the Pierre Auger Observatory (PAO) show departures from isotropy at the highest CR energies: a dipole above 8EeV \citep{pierre_auger_collaboration_observation_2017} and an intermediate scale indication of anisotropy at higher energies \citep{pierre_auger_collaboration_indication_2018}. 

One of the best candidate UHECR acceleration mechanisms is diffusive shock acceleration \citep[DSA; ][]{krymskii_regular_1977,axford_acceleration_1977,bell_acceleration_1978,blandford_particle_1978}, since it produces a well-defined spectral index in the particle momentum distribution and the acceleration timescale is generally shorter than second-order Fermi-type acceleration processes \citep[e.g.][]{blandford_particle_1987}. In supernova remnants, DSA provides a well-motivated framework for explaining CR acceleration up to $\sim100$TeV energies and possibly to the knee \citep[e.g.][]{bell_particle_2014}, while the acceleration of electrons up to $\sim$TeV energies in radio galaxy hotspots is also well explained by DSA theory \citep[e.g.][]{araudo_maximum_2018}. However, the situation at ultrahigh energies is different and the theoretical challenge of accelerating particles to 100 EeV and beyond is a significant one \citep[e.g.][]{blandford_acceleration_2000,lemoine2009}.

In this conference proceeding, we first review the requirements a source must meet for acceleration to ultrahigh energy, before presenting the case for radio galaxies as UHECR accelerators. Although we focus mainly on shock acceleration, many of the arguments are based on energetic constraints and apply regardless of the detailed acceleration physics. Much of this material has been presented by Matthews et al. \cite[][hereafter M18a, M18b]{M18a,M18b}, although the discussion here offers a briefer and more general perspective. The second paper should be consulted for more detail on the simulations and numerical method.

\section{Physical Requirements for UHECR Acceleration}
\label{sec:requirements}
% \subsection{The Hillas energy}
{\bf The Hillas energy:}
The characteristic maximum energy of a cosmic ray is set by the Hillas energy \citep{hillas_origin_1984}, given by
\begin{equation}\label{E-Hillas}
E_H = 0.9~\mathrm{EeV}~Z
\left( \frac{B}{\mu \mathrm{G}} \right)
\left( \frac{u}{c} \right)
\left( \frac{r}{\mathrm{kpc}} \right)
,
\end{equation}
where $B$ is the magnetic field, $u$ is the characteristic velocity (e.g. the shock velocity), $Z$ is the atomic number of the nucleus and $r$ is the size of the accelerator. We have neglected relativistic beaming, which will be a small effect in radio galaxies. The Hillas energy can be understood in terms of moving a particle of charge $Ze$ through an optimally correlated $-\boldsymbol{u}\times \boldsymbol{B}$ electric field, and is normally thought of as a characteristic maximum energy. Indeed, specific conditions need to be met to allow it to be reached (see section~\ref{sec:advantages}). 

% \subsection{The power requirement:}
{\bf The power requirement:}
If we consider a non-relativistic flow of velocity $u$ being focused through a cross-sectional area $r^2$, the kinetic power of this flow is $Q_k \sim \rho u r^2$. The magnetic field energy density is $B^2/2\mu_0$, so the {\em magnetic power} is then $Q_B \sim u r^2 B^2/2\mu_0$. We assume that the magnetic power is some fraction $\eta$ of the kinetic power, which in the context of diffusive shock acceleration can be thought of as an efficiency of magnetic field amplification at the shock. Combining all this with equation (1) allows us to write a minimum kinetic power requirement for acceleration to energy $E_H$, which is given by 
\begin{equation}\label{maxpower}
\frac{Q_0}{\mathrm{erg~s^{-1}}} = 10^{44}~Z^{-2}
\left( \frac{\eta}{0.1} \right)^{-1}
\left( \frac{E_H}{10 \mathrm{EeV}} \right)^2
\left( \frac{u}{0.1c} \right)^ {-1}
% \left( \frac{\Gamma}{1} \right)^ {-2}
,
\end{equation}
showing that a high kinetic power is required for UHECR acceleration. This power requirement is similar in nature to the Hillas energy and was proposed by a number of authors in different contexts \citep{lovelace_dynamo_1976,waxman_cosmological_1995,blandford_acceleration_2000,lemoine2009} and can include an additional beaming factor of $\Gamma^2$ \citep{nizamov_2011}. It is sometimes referred to as the Hillas-Lovelace-Waxman limit.

% \subsection{The non-relativistic requirement:}
{\bf The non-relativistic requirement:}
So far we have made arguments purely on energetic grounds, without considering the physical process for acceleration. It has been shown by various authors that acceleration to ultrahigh energies is problematic in relativistic shocks \cite{lemoine_electromagnetic_2010,reville_maximum_2014,bell_cosmic-ray_2018}. The reason for this is threefold: (i) Particles accelerated at relativistic shocks have steep spectra, so there is less energy available to drive turbulence on large (Larmor radii of EeV particles) scales; (ii) Relativistic shocks are quasi-perpendicular, so the UHECRs have to amplify the magnetic field within one Larmor radius of the shock else UHECRs are simply advected downstream; (iii) The UHECRs do not penetrate far downstream of the shock, and so there is very little time for the field to be amplified and stretch to the Larmor radius of EeV particles. Bell et al. \cite{bell_cosmic-ray_2018} consider ultra-relativistic shocks and find maximum energies on the order of 1 TeV, a prediction that matches constraints from the synchrotron cut-off in the hotspots of FRII radio galaxies \cite{araudo_evidence_2016,araudo_maximum_2018}. However, the exact shock velocity required for acceleration to ultrahigh energy is not clear; it must be significantly less than $c$, but not so small that it becomes restrictive in the Hillas energy or power requirement. 

\section{Particle Acceleration in Jet Backflows}
Given that relativistic shocks are generally poor UHECR accelerators, we therefore ask:
{\em are there non-relativistic shocks that form in radio galaxies that might offer more conducive conditions for UHECR acceleration?}

In addition to the termination shock, radio galaxies can also produce reconfinement shocks along the jet and drive a bow shock into the surrounding medium. The bow shock is slow and observational evidence from Centaurus A suggests it is not a good UHECR accelerator \cite{croston_high-energy_2009}. The  reconfinement shocks along the jet are interesting as particle accelerators and merit further investigation, although some of the issues with relativistic shocks still apply. Here, we instead focus on shocks that form after the jet material has passed through the relativistic termination shock. 

\subsection{Bernoulli's principle in jet lobes}
\begin{figure}[h]
\centering
   \includegraphics[width=\linewidth]{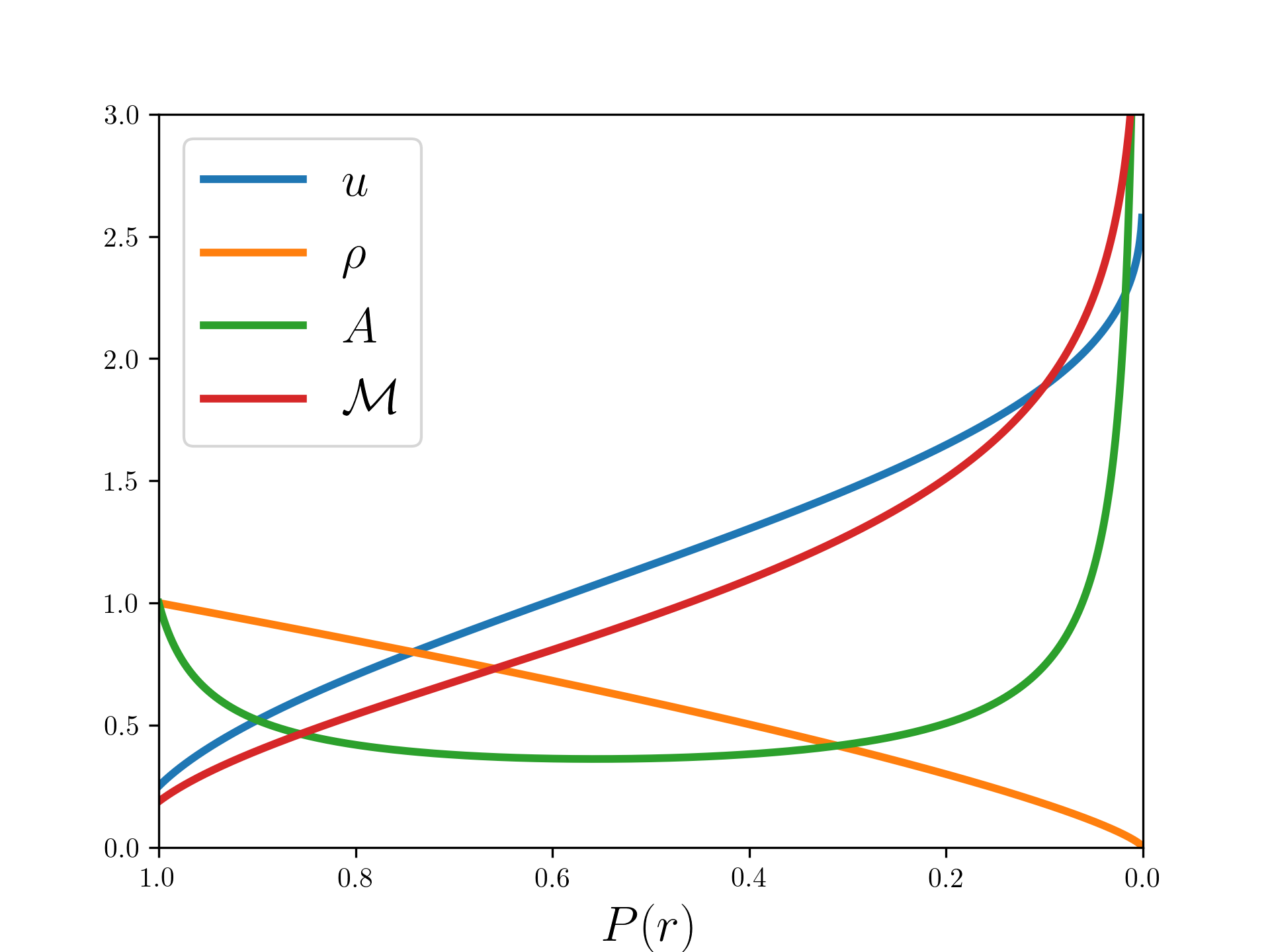}
\caption{The variation of physical quantities along a flux tube for a simple 1D model, illustrating how a drop in confining pressure can cause a flow to become supersonic. In this illustrative example, the Mach number ${\cal M}$ becomes greater than 1 when $P \lesssim 0.57$. Once a shock forms, the treatment breaks down because the adiabatic invariance no longer holds.
}
\label{fig-1}       % Give a unique label
\end{figure}

Neglecting relativistic effects, gravity and magnetic fields, the Bernoulli constant along a streamline of steady flow with adiabatic index $\gamma$ is
\begin{equation}
    \chi = \frac{u^2}{2} + \frac{\gamma}{\gamma-1}\frac{P}{\rho} = {\rm constant}. 
\end{equation}
We consider a flux tube of steady flow with the initial condition set by the strong shock jump condition. We also make use of the adiabatic constraint, $P \rho^{-\gamma} =  {\rm constant}$, and mass conservation, $\rho u A = {\rm constant}$. Under these assumptions, we can evaluate $\rho$, $u$, $A$ and the Mach number, ${\cal M}$, as a function of pressure. The resulting profiles are shown in Fig.~\ref{fig-1}, with the units normalised so that, initially, $P=1, \rho=1, A=1, u=0.25$. The flow becomes supersonic once $P \lesssim 0.57$ - after this point, the only way the flow can slow down is via a shock. 

\begin{figure}[h]
\centering
   \includegraphics[width=0.8\linewidth]{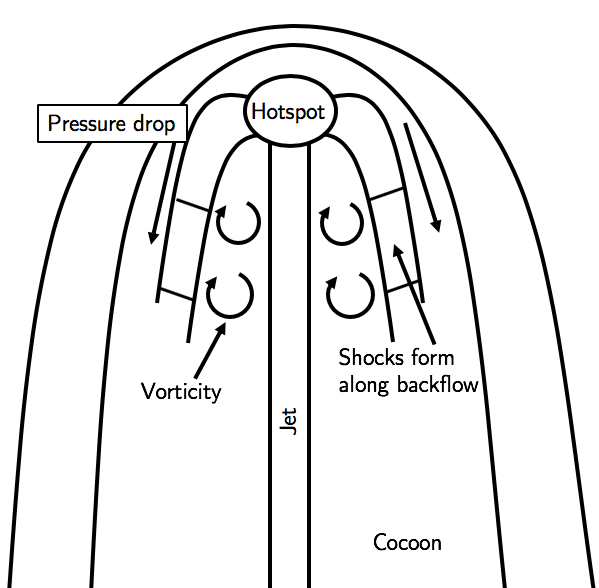}
\caption{
A schematic of the jet, hotspot, backflow and cocoon in a radio galaxy.
}
\label{fig-2}       % Give a unique label
\end{figure}

Radio galaxies can have large pressure gradients between their hotspots and cocoons. The plasma from the hotspot is therefore funneled backwards away from the hotspot, forming narrow streams closely analogous to our flux tube model (see schematic in Fig.~\ref{fig-2}). The streams come into pressure equilibrium with the surroundings, so the profiles shown in Fig.~\ref{fig-1} become an approximate representation of the flow along a streamline. This analysis is valid until a shock is formed. The Mach number of the shock will depend on the pressure profile along the stream. 

\begin{figure*}[h]
\centering
   \includegraphics[width=0.9\linewidth]{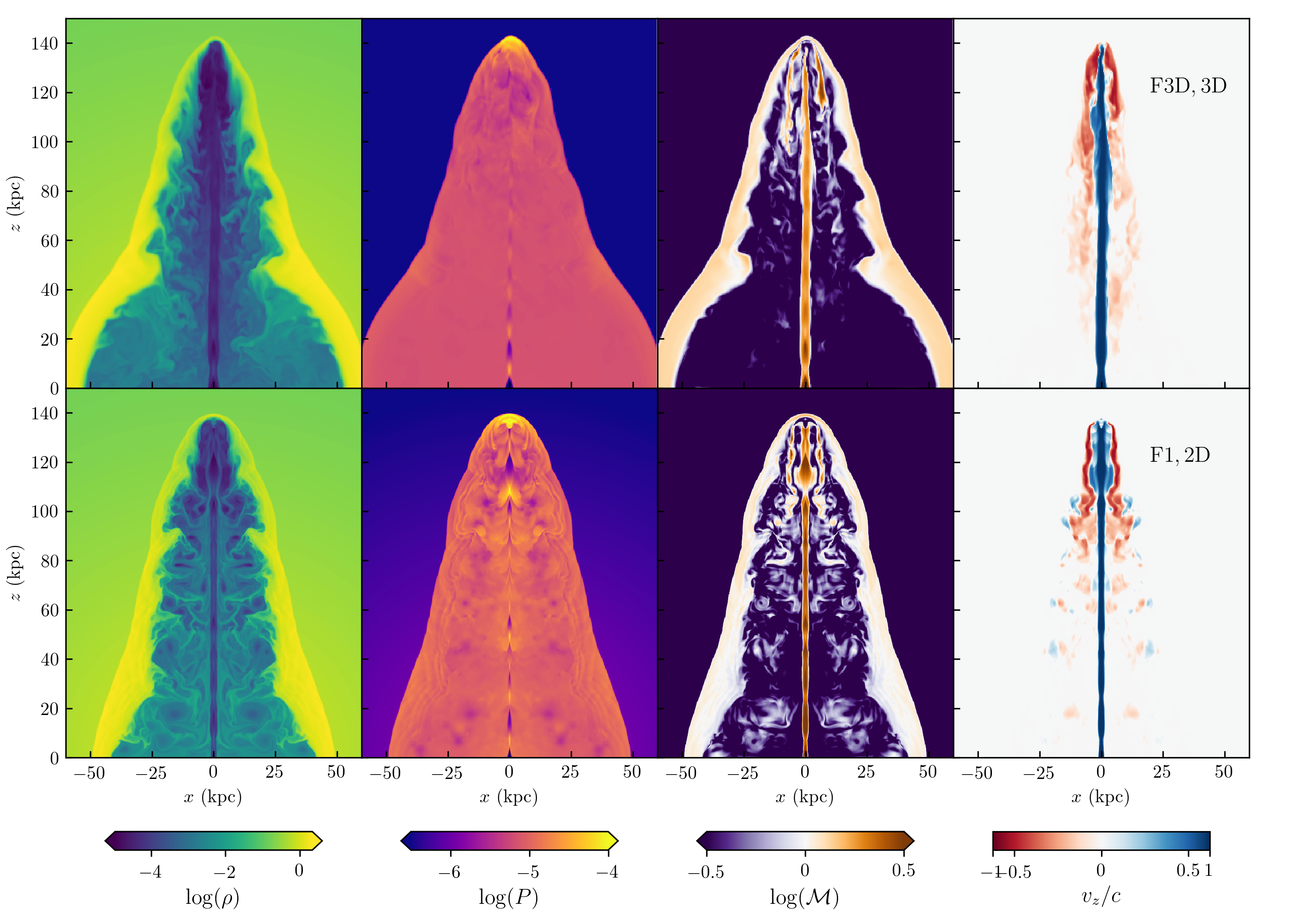}
\caption{Four physical quantities (density, pressure, Mach number and vertical velocity) in the jet shown as a slice at $y=0$~kpc, where the domain extends from $-60$~kpc to $60$~kpc in $x$ and $y$ (the 2D simulation has been reflected about the $z$ axis). The jet is supersonic and produces a termination shock. The high-pressure post-shock material inflates a low-density cocoon and drives a bow shock into the surrounding material. As predicted by the Bernoulli analysis, fast backflows can form between the jet head and cocoon. These backflows are supersonic and produce shocks. 
}
\label{fig-3}       % Give a unique label
\end{figure*}

\subsection{Hydrodynamic Simulations}
The treatment outlined above is clearly a simplification, so we turn to hydrodynamic simulations to further explore shocks in backflows. The relativistic hydrodynamic (RHD) simulations are described in detail by M18b. We use \textsc{PLUTO} \citep{pluto} to solve the equations of RHD. We inject a smoothed top-hat jet into a cluster with density and pressure profiles chosen to roughly match typical profiles from Ineson et al. \cite{ineson_link_2015}. The cluster is initially in hydrostatic equilibrium with small density perturbations ($\delta \rho / \rho \sim10^{-10}$), and we have verified that decreasing the magnitude of the cluster pressure and gravitational field by a factor of $10^7$ does not significantly alter the behaviour of the jet, particularly outside the cluster core. We present two simulations, a 3D simulation (F3D) with jet power $10^{45}$~erg~s$^{-1}$ and a 2D cylindrical simulation (F1) with jet power $2.69\times10^{45}$~erg~s$^{-1}$. Both jets are light with respect to their surroundings, with density contrasts ($\eta=\rho_j/\rho_0$) of $1.88\times10^{-5}$ (F3D) and $9.71\times10^{-5}$ (F1), which corresponds to relativistic generalisations of the density contrast of $1.92\times10^{-4}$ and $10^{-3}$, respectively. The full simulation parameters are given in table 1 of M18b.

The simulation results are presented in Fig.~\ref{fig-3}, where we show slices of density, pressure, Mach number and $z$ velocity in the $x-z$ plane for a snapshot of the F1 and F3D simulations. The jet produces reconfinement shocks along its path, and forms a classic two-shock structure. It inflates a low-density cavity which is filled by rapidly backflowing plasma from the jet head. These backflows can become supersonic, qualitatively validating the results of the 1D Bernoulli argument, and can produce non-relativistic shocks. These shocks can accelerate particles via DSA. 

\subsection{Advantages of backflows as ultrahigh energy particle accelerators}
\label{sec:advantages}
Shocks in backflows have a number of interesting properties as UHECR accelerators. The Hillas energy gives a constraint on the shock properties that is necessary but not sufficient, as shown by e.g. \citep{lagage_cosmic-ray_1983, bell_cosmic-ray_2013}. At a single shock, the Hillas energy only applies in the limit of Bohm diffusion, i.e. $\lambda \sim r_g$, where $\lambda$ is the scattering mean-free path and $r_g$ is the Larmor radius. The maximum CR energy is therefore limited by the time available to generate turbulence on the scale of $r_g$. In backflows, and other complex, turbulent flows, the turbulence is not only self-generated by CRs. Field lines can be stretched and distorted by the turbulent flow close to the jet head, while passing through multiple shocks offers more than one opportunity for CR-generated turbulence. The seed field for this turbulence is the magnetic field at the hotspot, which can already be relatively large due to amplification at the termination shock \citep{araudo_maximum_2018}. Furthermore, the shock velocities can span a range of values. The combination of these factors means that many of the problems associated with acceleration at relativistic shocks can be avoided. Detailed calculations are needed to validate these statements, but generally we expect shocks in the backflow to produce CRs with energies approaching Hillas. 

\subsection{Maximum energy estimate}
To estimate the maximum CR energy in the simulations, we need an estimate of the characteristic shock size, magnetic field, Mach number and velocity. We used Lagrangian tracer particles to track the passage of a fluid element. Shocks were identified by a pressure jump, $\Delta P/P > 0.2$ and requiring that $\nabla \cdot \boldsymbol{u} < 0$. Shock properties were recorded each time the particle crossed a shock, which also allowed the number of shock crossings to be recorded. The shock detection procedure is described in more detail by \cite{M18b}. We also used a DBSCAN clustering analysis \cite{ester_density-based_1996} to detect shock structures on the Eulerian grid in order to measure shock size. We restrict the tracer particle analysis to the 2D simulation, finding approximately $10\%$ of particles passed through a shock with ${\cal M} >3$. The shocks had characteristic sizes and velocities of $r_s\approx2$kpc and $u_s\approx0.2c$, respectively. We estimated the magnetic field as a fraction of the total energy density at the shock, i.e. $\bar{B} = [2 \eta \mu_0 (\rho u^2 + U_{\mathrm{th}}) ]^{1/2}$
obtaining values up to a maximum of $150\mu$G for a departure from equipartition, $\eta$, of 0.1. Taking these characteristic values and substituting into equation (1), we obtain 
\begin{equation}\label{E-Hillas}
E_\mathrm{max} \approx 50~\mathrm{EeV}~Z
\left( \frac{B}{140 \mu \mathrm{G}} \right)
\left( \frac{u_s}{0.2c} \right)
\left( \frac{r_s}{2\mathrm{kpc}} \right).
\end{equation}
This maximum energy can be higher by approximately a factor $N$ if a particle passes through $N$ shocks on its passage through the backflow. This estimate indicates that shocks in the backflows of radio galaxies are plausible UHECR acceleration sites, particular if the UHECRs are composed of mostly heavy nuclei beyond $\sim 10$~EeV.

\section{Power requirements and UHECR Escape}
\label{sec:escape}
We have demonstrated that shocks in the lobes of radio galaxies can have a range of shock velocities, with Hillas energies of up to $\sim5\times10^{19}$eV. As a result, the scenario of UHECR acceleration in secondary shocks in radio galaxies meets two of our three minimal requirements from section~\ref{sec:requirements}. The power of the jet is an input parameter to the simulations, designed to mimic typical FRII jet powers of $\sim10^{45}$~erg~s$^{-1}$ \citep{ghisellini_2001}, so we must turn to observational constraints to ascertain if powerful radio galaxies are common and local enough to be plausible UHECR sources.

Using empirical relationships between jet kinetic power and radio luminosity \citep[e.g.][]{cavagnolo_relationship_2010}, it is possible to integrate over the radio galaxy luminosity function using the minimum power requirement as a lower limit. Such an exercise reveals that, on average, radio galaxies are common and energetic enough to reproduce the observed UHECR luminosity at Earth (M18b). A more detailed analysis of UHECRs from radio galaxies, which considers the spectrum produces as well as propagation effects, has been carried out by Eichmann et al. \citep{eichmann_ultra-high-energy_2018}. They indeed find that radio galaxies are plausible UHECR sources. However, there are very few powerful radio galaxies within a canonical GZK horizon of $\sim100$Mpc that meet the power requirement from section~\ref{sec:requirements} \citep{massaglia_role_2007,eichmann_ultra-high-energy_2018,M18a}. It is therefore natural to explore whether the jet powers in local radio galaxies could have been higher in the past.

In M18a, we suggested that the giant lobes of two of the nearest radio galaxies -- Centaurus A and Fornax A -- could have been produced during more powerful past outbursts, particularly in the case of Fornax A. The lobes of both sources are 100s of kpc across \cite{ekers_large-scale_1983} with extremely large energy contents of $\sim5\times10^{58}$~erg \cite{lanz_constraining_2010} in Fornax A and $\sim10^{59-60}$~erg \citep{wykes_mass_2013,eilek_dynamic_2014} in Centaurus A. This total energy content is large compared to the current jet powers (M18a). The approximate BH masses of the sources are $10^{8}~M_\odot$ for Fornax A \citep{nowak_supermassive_2008} and $5\times10^{7}~M_\odot$ for Centaurus A \citep{cappellari_mass_2009}. Thus, based on the Eddington luminosity, outbursts with jet powers of $>10^{45}$~erg~s$^{-1}$ are feasible, which would allow the power requirement from section~\ref{sec:requirements} to be met. Furthermore, both sources are extended gamma-ray sources, and Fornax A requires a hadronic component in order to fit the \textit{Fermi} data \citep{ackermann_fermi_2016}, implying that it is indeed a ``reservoir'' for CRs, albeit at much lower energy. If this scenario -- one of giant-lobed radio galaxies acting as slowly-leaking reservoirs of UHECRs accelerated during past outbursts -- is feasible, then the lobes must be able to confine the UHECRs for a reasonably long time, but not so long that they are destroyed by interaction losses.

We can estimate the escape time for a CR from the lobes by assuming that they diffuse out of the lobe with a coefficient $D$. The escape time for a CR of rigidity $E/Z$ from a lobe of size $R$ is then approximately given by
\begin{equation}
    t_{\mathrm{esc}} \sim 9~\mathrm{Myr}
    \left( \frac{D}{D_\mathrm{B}} \right)^{-1}
    \left( \frac{B}{\mu \mathrm{G}} \right)
    \left( \frac{E/Z}{10 \mathrm{EV}} \right)^{-1}
    \left( \frac{R}{100 \mathrm{kpc}} \right)^2
\end{equation}
where $D_B=r_g c /3$ is the Bohm diffusion coefficient. The equation above will be modified depending on exactly how the CRs escape from the lobes; for example, if the CRs undergo a $\nabla B$ drift or stream along field lines. Alternatively, local magnetic field structures larger than $r_g$ may effectively confine the particles for even longer times. Other authors have considered the delayed escape of UHECRs from clusters \cite{rordorf2004,kotera2009}, with Kotera et al. \cite{kotera2009} finding CRs with rigidity $10$EV can be confined for $\sim10$ Myr in a $\approx10\mu$G field. In these simulations, the escaping spectrum changes over time, and a hard spectral index may be possible if lower energy CRs are killed off by interaction losses before they can escape. Generally, confinement in the dormant radio lobe for $\gtrsim$Myr timescales is likely, but this needs qualification and more detailed calculation.

\section{UHECR Arrival Directions}
In the past few years, the first evidence for departures from isotropy at ultrahigh CR energies has emerged. TA data shows a `hotspot' above 57 EeV \cite{abbasi_indications_2014}, while the PAO data shows a dipole above 8 EeV \cite{pierre_auger_collaboration_observation_2017} and indications of anisotropy on intermediate scales at higher energies \citep[][hereafter A18]{pierre_auger_collaboration_indication_2018}. 

Can radio galaxies explain the PAO data? The A18 study finds a $4\sigma$ correlation with the positions of starburst galaxies (SBGs) and a $2.7\sigma$ correlation with gamma-ray AGN from the second catalog of hard \textit{Fermi}-LAT sources (2FHL, \citep{ackermann_2fhl:_2016}). The residuals from the AGN fit show  excesses at  $l = 308^\circ,b = 26$ and $l = 275^\circ,b = -75$ (referred to as HS1 and HS2 by M18a). The 3FHL data has been recently made available \citep{ajello_3fhl:_2017} and, unlike 2FHL, includes extended gamma-ray sources. As a result it includes Fornax A, but 2FHL does not. The 3FHL catalogue also reports a factor $2.25$ increased flux for Centaurus A. 

Both including Fornax A and an increased flux from Centaurus A might be expected to increase the significance of the Auger result, but only if magnetic deflection is able to account for the $22.5^\circ$ offset between Fornax A and the excess at southern Galactic latitude. Indeed, simply including Fornax A in the PAO analysis does not improve the deviation from isotropy (Biteau, private communication on behalf of the Pierre Auger Collaboration). In this analysis the luminosity proxy is set by the 2FHL gamma-ray luminosity and the search radius is a free parameter that is optimised in the fit. It might be interesting in future studies to modify the assumptions of the analysis. For example, one could envision making the search radius dependent on source distance, $d$, as might be expected if the deflection is extragalactic (e.g., $\delta \theta \sim (l_c d)^{1/2}~E^{-1}~Z B_{\mathrm{EG}}$, where $B_{\mathrm{EG}}$ and $l_c$ are the strength and coherence length of the extragalactic magnetic field \citep{sigl_ultrahigh_2003,eichmann_ultra-high-energy_2018}). However, it is extremely difficult to account for magnetic deflection in a statistically rigorous and physically motivated way. 

The effective attenuation length and luminosity proxies that go into the analysis are also important. The effective attenuation length is strongly composition dependent \citep[e.g.][]{alves_batista_cosmic_2015,wykes_uhecr_2017} and is obtained from a combined fit to the PAO spectrum above $10^{18}$eV and $X_{\mathrm{max}}$ distributions \citep{pierre_auger_collaboration_combined_2017}. In this combined fit, the source population is assumed to be homogenous and isotropic, although the sensitivity to source evolution with redshift is explored. The favoured model (Scenario A in A18) has a hard spectrum ($k=0.96$). As a result, the model has a relatively short attenuation length and low maximum rigidity ($\log R_{\mathrm{max}}=18.78$).
% and the flux suppression is caused by a combination of the maximum rigity at the sources and propagation effects.
 It is possible to fit the data with softer spectral indices (e.g. $k=2.04$), but the fit to the $X_{\mathrm{max}}$ data is strongly disfavoured. The luminosity proxy is also problematic, since the radiative luminosities in the \textit{Fermi} bands and at radio wavelengths probe energy regimes orders of magnitude lower than EeV energies. Furthermore, the UHECRs can experience delays with respect to radiative signatures 
if they are confined in a lobe or cluster (see section~\ref{sec:escape}).

\begin{figure}[h]
\centering
   \includegraphics[width=\linewidth]{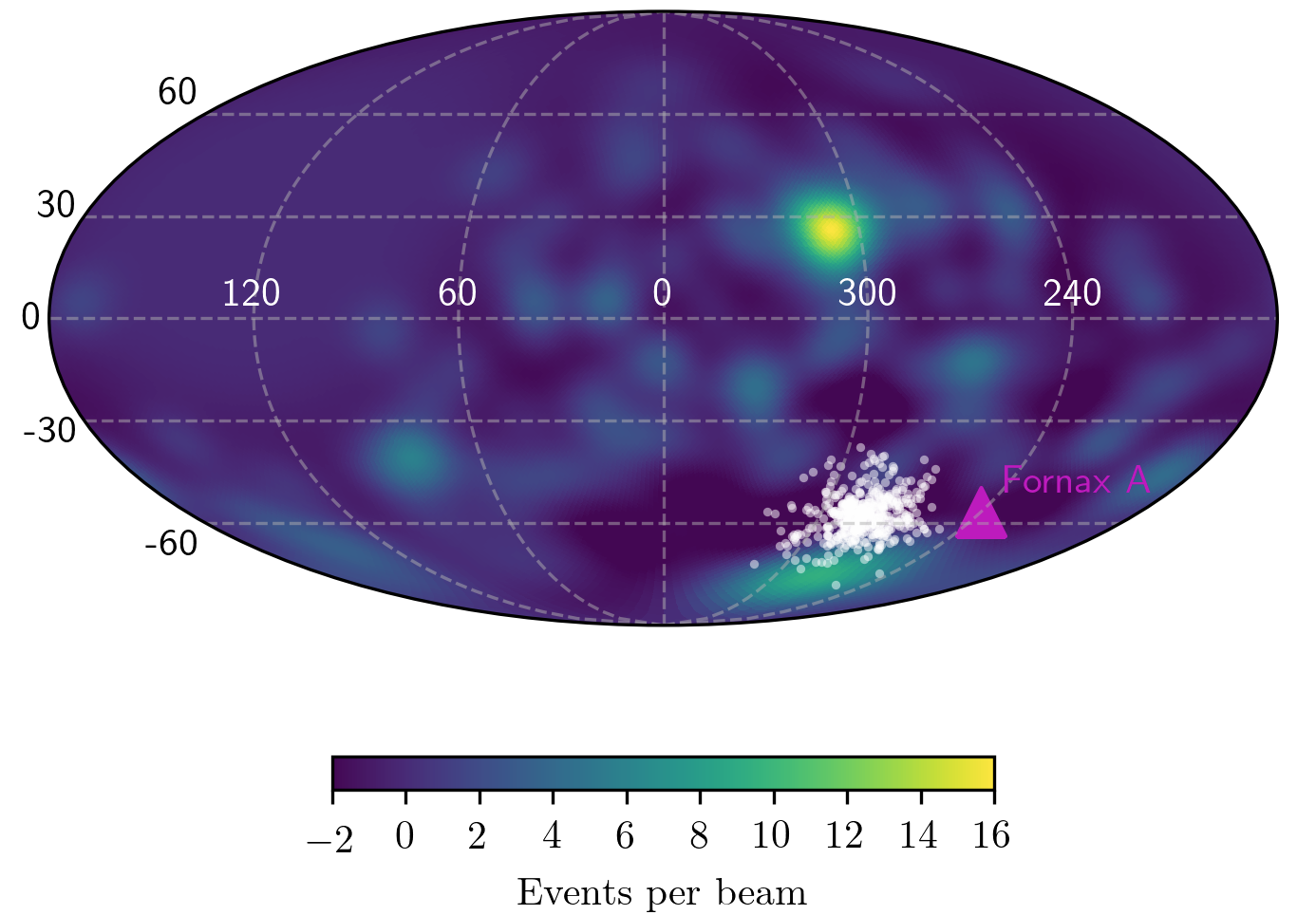}
\caption{The observed excess map from A18 in Galactic coordinates for events with energies $>60$~EeV, with the position of Fornax A overlaid. The projection is the same as A18 and the formulae are given in M18a. The white dots show the simulated directions of $10$~EV CRs propagated from Fornax A through the Jansson \& Farrah \citep{jansson_new_2012} Galactic magnetic field model as described in the text. 
}
\label{fig-5}       % Give a unique label
\end{figure}

Given the uncertainties described above, we cannot hope at this stage to conclusively associate a given source population with UHECRs based purely on anisotropies. We instead make some plausibility arguments. We show the observed excess map for energies $>60$~EeV from A18 in Galactic co-ordinates in Fig.~~\ref{fig-5}. Overlaid on the map is the position of Fornax A and the results of a CRPropa 3 \citep{alves_batista_crpropa_2016} simulation. In this simulation, we propagate $E/Z=10$EV particles with initial positions drawn from a Fisher distribution centred on Fornax A with concentration parameter $\kappa=51.26$, equivalent to a spread of $11.3^\circ$ due to a turbulent extragalactic field. The Jansson \& Farrah \citep{jansson_new_2012} Galactic magnetic field model is then applied using the inbuilt lenses in CRPropa 3. The direction of deflection is roughly towards the excess, although it cannot be considered a good fit to the data. The dispersion of the points, offset from the hotspot and deflection from Fornax A are all of comparable magnitude. For lower rigidities ($E/Z$), the deflection curves the directions towards the Galactic plane, while for higher rigidities the deflection is small
\citep{smida_ultra-high_2015}. The magnitude of the deflection required for Fornax A to explain the HS2 excess is reasonable, but the direction using one specific model is not correct. These conclusions depend on the systematic uncertainties in the magnetic field model, which may be large \citep[e.g.][]{unger_uncertainties_2017}. 

\section{Discussion \& Conclusions}
\label{sec:conc}
We first described three requirements UHECR sources must satisfy; they must meet the Hillas condition, supply enough power and produce non-relativistic shocks. We then summarised the results of two papers, showing that (i) backflow shocks in the lobes of radio galaxies can meet the required criteria, and (ii) Fornax A and Centaurus A make for compelling UHECR sources. In order for these ideas to be developed further, future work could include: (i) studying CR escape from radio lobes; (ii) searching for observational evidence of shocks in backflows; (iii) modelling magnetic field amplification and particle acceleration along a backflowing stream; and (iv) varying the luminosity proxies or, if possible, allowing for Galactic magnetic field deflections in the UHECR anisotropy analysis. Combined observational results from TA and PAO will also be crucial for identifying UHECR accelerators during this exciting time for cosmic ray astrophysics. 

\section*{Acknowledgements}
We thank the anonymous referee  and the organisers and attendees of UHECR 2018, as well as J. Biteau, A. Watson, B. Eichmann and A. Taylor for helpful discussions. This work is supported by STFC grant ST/N000919/1. A.T.A. thanks the Czech Science Foundation (ref. 14-37086G) in Prague, and the EU COST Action (ref. CA16104). We acknowledge the use of matplotlib \cite{matplotlib}, healpy \cite{healpy}, scikit-learn \cite{scikit-learn} and astropy \cite{the_astropy_collaboration_astropy_2018}.

\end{document}